# Evaluating analytic gradients of pulse programs on quantum computers

Korbinian Kottmann[1], Nathan Killoran[1]

[1] Xanadu, Toronto, ON, M5G 2C8, Canada

**Abstract**

Parametrized pulse programs running on quantum hardware can be differentiated via the stochastic parameter-shift (SPS) rule. We overcome the intrinsically approximate nature of SPS by introducing a new analytic method for computing gradients of pulse programs, that we coin ODEgen [/oʊ-diː-dʒɛn/]. In this new method, we find effective generators of pulse gates using a differentiable ordinary differential equation (ODE) solver. These effective generators inform parameter-shift rules that can be evaluated on quantum hardware. We showcase simulated VQE examples with realistic superconducting transmon systems, for which we obtain lower energies with fewer quantum resources using ODEgen over SPS. We further demonstrate a pulse VQE run with gradients computed via ODEgen entirely on quantum hardware.

## 1. Intro

Contemporary quantum computers based on superconducting qubits [1], ion trap systems [2], and neutral Rydberg atoms [3] are operated at their core through electromagnetic signals that drive the physical qubits. This low-level control is a necessity to run digital quantum circuits that are transpiled down to their calibrated pulses on hardware. Optimizing the calibration of digital quantum gates is an intricate step to achieving fault tolerance and, at the same time, improves the fidelities of noisy circuit executions without error correction. Pulse-level control can further be seen as a new avenue of controlling quantum computers more efficiently for noisy near-term applications [4, 5, 6].

The generalized parameter-shift rules [7, 8] enable *analytic* gradient computation of parametrized quantum circuits of the form $\mathcal{L} = \langle\psi(\theta)|H|\psi(\theta)\rangle$ with $|\psi(\theta)\rangle = \prod_j e^{-i\theta_j G_j} |0\rangle$ and Hermitian generators $G_j$. For pulse gates, where the operators generating the system's evolution may vary continuously in time, there is the stochastic parameter-shift rule that allows approximate gradient estimation via Monto Carlo integration [9, 10] (see Sec. 2.1).

In this paper, we introduce a new *analytic* method to compute gradients of pulse programs in analogy to recently introduced methods to compute gradients of general `SU(N)` gates [11]. We leverage ordinary differential equation (ODE) solvers to obtain so-called *effective generators* of pulse gates. These are then decomposed and used to construct analytic parameter-shift rules that can be executed on quantum hardware. A brief summary of the procedure is illustrated in Fig. 1, with details in Sec. 2.2. We find that this has an advantage over the stochastic parameter-shift rule in relevant systems with comparable or lower quantum resources (i.e., number of computed expectation values).

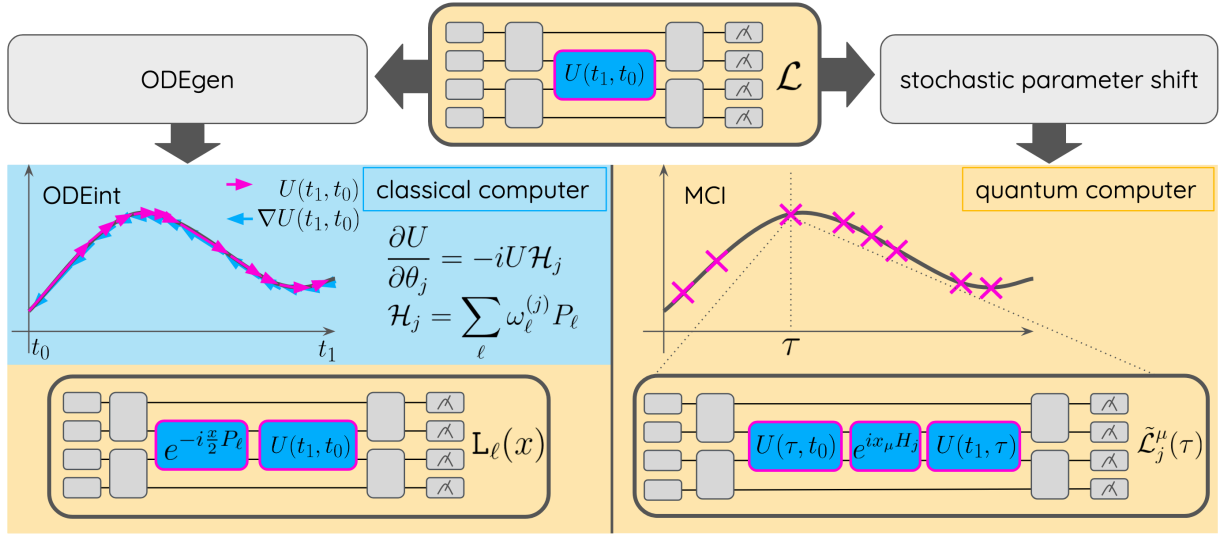

Fig. 1: ODEgen method for computing gradients of pulse programs, introduced in this paper, in comparison with the stochastic parameter-shift rule. We are considering a cost function $\mathcal{L}$ with a circuit that includes a pulse gate $U(t_1, t_0; \theta)$, whose $\theta$-dependence we drop for brevity. ODEgen avoids computing approximate integrals via Monte Carlo integration (MCI) by leveraging a differentiable ordinary differential equation solver (ODEint) with adaptive step sizes. (ODEgen) On a classical computer, we use ODEint to obtain both the pulse gate unitary $U(t_1, t_0)$ as well as its gradient $\nabla U(t_1, t_0)$. The differently sized arrows indicate the adaptive step sizes during the forward pass for integrating the ODE [12] and backward pass for gradient computation [13]. This way we can compute the effective generator $\mathcal{H}_j$, from which we can construct parameter-shift rules with shifted expectation values $\mathtt{L}_\ell(x)$, that can be executed on the quantum computer. The details are described in Sec. 2.2. (Stochastic parameter shift) Expectation values $\tilde{\mathcal{L}}_j^\mu(\tau)$ with shifted evolutions $U(t_1, \tau)e^{-ix_\mu H_j}U(\tau, t_0)$ are sampled uniformly at random intermediate times $\tau \in [t_0, t_1]$ to approximate the integral Eq. (12) via MCI. The details are described in Sec. 2.1.

Gradient-based quantum optimal control algorithms like the gradient ascent pulse engineering (GRAPE) [14, 15] algorithm and modern variations thereof [16, 17] retain a number of limitations. Often they are limited to run on small-scale quantum simulations, or with approximate gradients (e.g., finite differences), or they employ a restricted set of parametrizations. ODEgen enables us to run these algorithms natively on quantum hardware of arbitrary system size, as long as interactions induced by pulses are restricted to small subsystems. We can also obtain analytic gradients of arbitrary families of parametrizations for the pulses. A related class of algorithms, variational quantum algorithms [18, 19], and in particular the variational quantum eigensolver (VQE), can be run on quantum hardware using pulse programming [5, 10]. We see advantages of using ODEgen compared to SPS by finding lower energies in the VQE examples highlighted in [10] with fewer quantum resources. We further showcase a small-scale VQE pulse programming example on a superconducting quantum computer from *Oxford Quantum Circuits* with so-called coaxmon qubits [20] accessed through `PennyLane` [21] and Amazon Braket [22] in the cloud.

We recap pulse gates and their derivatives in Sec. 2, where we introduce ODEgen in Sec. 2.2 for analytic gradient computation. We then perform numerical experiments in Sec. 3 with transmon qubit Hamiltonians to showcase the advantages of ODEgen and, more generally, the utility of differentiable pulse programs for variational quantum algorithms and gate calibration.

## 2. Gradients of pulse programs

We refer to a *pulse sequence*, *pulse program*, *pulse gate* or simply just a *pulse* as the controlled time evolution of the quantum state encoded in a quantum computer due to some parametrized, time-dependent control terms $f_j(\theta, t) H_j$. These are typically realized by driving the physical qubits of the system with a controlled electromagnetic field (see Sec. 3 for a realistic model). In general, the time-dependent Hamiltonian can be written as

$$H(\theta, t) = H^{\text{drift}} + \sum_{j=1}^{N_g} f_j(\theta, t) H_j, \tag{1}$$

where we assign the drift term $H^{\text{drift}}$ the special role of a constant, uncontrolled contribution. There are $N_g$ drive generator terms $H_j$ parametrized by the scalar pulse envelopes $f_j$. The physical system evolves according to the time-dependent Schrödinger equation

$$\frac{d}{dt} |\psi(t)\rangle = -i H(\theta, t) |\psi(t)\rangle \tag{2}$$

from an initial state $|\psi(t_0)\rangle$ to a final state $|\psi(t_1)\rangle = U(t_1, t_0; \theta) |\psi(t_0)\rangle$, where the pulse gate realizes the unitary evolution $U(t_1, t_0; \theta)$.

We are interested in the derivative $\nabla_\theta U(t_1, t_0; \theta)$ and, for the sake of notational brevity, drop the explicit dependence on $\theta$. The unitary evolution can be written as a product of infinitesimal time steps $\delta_n = \frac{t_1 - t_0}{n}$,

$$U(t_1, t_0) = \lim_{n \to \infty} e^{-i\delta_n H(\theta, t_1)} e^{-i\delta_n H(\theta, \tau_{n-1})} \ldots e^{-i\delta_n H(\theta, \tau_1)} e^{-i\delta_n H(\theta, t_0)}, \tag{3}$$

with infinitesimal intermediate times $\tau_j = t_0 + j\delta_n$. Differentiating with respect to $\theta$ yields

$$\frac{\partial}{\partial \theta} U(t_1, t_0) = \lim_{n \to \infty} \sum_{j=0}^{n} \left( \ldots \exp(-i\delta_n H(\theta, \tau_{j+1})) \frac{\partial [-i\delta_n H(\theta, \tau_j)]}{\partial \theta} \exp(-i\delta_n H(\theta, \tau_j)) \ldots \right). \tag{4}$$

Together with the identity $U(t_1, \tau) U(\tau, t_0) = U(t_1, t_0)$ and identifying the Riemann sum, we can write the derivative as an integral

$$\frac{\partial}{\partial \theta} U(t_1, t_0) = -i \int_{t_0}^{t_1} d\tau \, U(t_1, \tau) \frac{\partial H(\theta, \tau)}{\partial \theta} U(\tau, t_0). \tag{5}$$

Intuitively, we can recognize the above equation as a kind of chain rule at each point $\tau$, integrated across the entire domain $[t_0, t_1]$.

### 2.1. Stochastic parameter-shift rule for pulses

We are interested in computing the gradient of an objective function $\mathcal{L}(\theta) = \langle \psi_0 | U^\dagger(\theta) H^{\text{obj}} U(\theta) | \psi_0 \rangle$ that computes the expectation value of an objective Hamiltonian $H^{\text{obj}}$ (e.g., a molecular Hamiltonian or the Hamiltonian of a spin system). For a circuit composed solely of what we call *digital gates* $U(\theta) = \prod_j e^{-i\theta_j G_j}$ with some Hermitian generators $G_j$, we can compute *analytic* gradients using the general parameter-shift rule

$$\frac{\partial \mathcal{L}}{\partial \theta_j} = \sum_{\mu=1}^{2R} y_\mu \mathcal{L}(\theta_j^\mu) \tag{6}$$

with coefficients $\{y_\mu\}$, shifted parameters $\{\theta_j^\mu\}$ and unique eigenvalue differences $R$; see [8] for details. Note that for Pauli-word generators with eigenvalues $\pm 1$, this is just the common two-term parameter-shift rule

$$\frac{\partial \mathcal{L}}{\partial \theta_j} = \frac{1}{2}\left(\mathcal{L}\left(\boldsymbol{\theta} + \boldsymbol{e}_j \frac{\pi}{2}\right) - \mathcal{L}\left(\boldsymbol{\theta} - \boldsymbol{e}_j \frac{\pi}{2}\right)\right), \tag{7}$$

where the $j$-th unit vector $\boldsymbol{e}_j$ is used to shift the $j$-th component of $\boldsymbol{\theta}$.

For the case in which the state evolves according to a time-dependent pulse sequence,

$$\mathcal{L} = \langle \psi_0 | U(t_1, t_0)^\dagger H^{\text{obj}} U(t_1, t_0) | \psi_0 \rangle, \tag{8}$$

the only solution previously available was the stochastic parameter-shift (SPS) rule [10]

$$\frac{\partial \mathcal{L}}{\partial \boldsymbol{\theta}} = \int_{t_0}^{t_1} d\tau \sum_j I_j(\tau) \tag{9}$$

with the integrands

$$I_j(\tau) = \frac{\partial f_j(\tau)}{\partial \boldsymbol{\theta}} \sum_{\mu=1}^{2R} y_\mu \tilde{\mathcal{L}}_j^\mu(\boldsymbol{\theta}, \tau). \tag{10}$$

$\tilde{\mathcal{L}}_j^\mu(\boldsymbol{\theta}, \tau)$ are expectation values $\langle \psi_j^\mu(\tau) | H^{\text{obj}} | \psi_j^\mu(\tau) \rangle$ under shifted evolutions $|\psi_j^\mu(\tau)\rangle = U(t_1, \tau) e^{-ix_\mu H_j} U(\tau, t_0) |\psi_0\rangle$. The coefficients $\{y_\mu\}$ and shifts $\{x_\mu\}$ are determined by the pulse generators $H_j$ according to the general parameter-shift rules [8]. In the case of single-qubit Pauli operators, we simply have

$$\frac{\partial \tilde{\mathcal{L}}}{\partial \boldsymbol{\theta}} = \int_{t_0}^{t_1} d\tau \sum_j \frac{\partial f_j(\tau)}{\partial \boldsymbol{\theta}} \left(\tilde{\mathcal{L}}_j^+(\tau) - \tilde{\mathcal{L}}_j^-(\tau)\right), \tag{11}$$

with $\tilde{\mathcal{L}}_j^\pm(\tau) = \langle \psi_j^\pm | H^{\text{obj}} | \psi_j^\pm \rangle$ and $|\psi_j^\pm\rangle = U(t_1, \tau) e^{-i(\pm \frac{\pi}{4}) H_j} U(\tau, t_0) |\psi_0\rangle$. For the case that the loss function Eq. (8) consists of more than just the time-dependent pulse program, e.g., $V_1 U(t_1, t_0) V_0$, one simply subsitutes $|\psi_j^\pm\rangle = V_1 U(t_1, \tau) e^{-i(\pm \frac{\pi}{4}) H_j} U(\tau, t_0) V_0 |\psi_0\rangle$.

In order to evaluate this integral on quantum hardware, it can be approximated via Monte Carlo sampling

$$\int_{t_0}^{t_1} d\tau I(\tau) \approx \frac{t_1 - t_0}{N_s} \sum_{\tau \sim \mathcal{U}[t_0, t_1]} I(\tau) \tag{12}$$

by drawing $N_s$ random samples from the uniform distribution on the given time interval, $\mathcal{U}[t_0, t_1]$.

The number of expectation-value evaluations on the quantum computer, i.e., the quantum resources, needed to estimate the gradient via the SPS rule is given by

$$\mathcal{R}_{\text{SPS}} = N_s N_g 2R, \tag{13}$$

where $N_s$ is the number of samples used in the Monte Carlo integration in Eq. (12) and $N_g$ is the number of drive generators $H_j$. According to the central limit theorem, the error or standard deviation of the gradient estimate scales as $\frac{1}{\sqrt{N_s}}$.

## 2.2. Analytic gradients of pulse programs using `ODEgen`

In this paper, we introduce an alternative method for computing gradients of pulse programs on quantum hardware that we coin *ODEgen*. It is a generalization of the `SU(N)` gradients introduced in [11] to pulse programs. The big advantage of the method is that it is analytic and does not involve approximating an integral like in Eq. (12). Instead, we compute so-called effective generators, from which we can construct parameter-shift rules that we can execute on quantum hardware. We obtain these effective generators classically by leveraging a just-in-time compiled and differentiable ODE solver with adaptive step sizes in `jax` [23]. ODE solvers with built-in automatic differentiation have recently risen to prominence in machine learning, due in part to concepts such as Neural ODEs [12, 13]. These solvers can compute both $U(t_1, t_0)$ and its gradient with respect to trainable parameters $\nabla U(t_1, t_0)$ using a combination of forward/backward passes. The overall idea is illustrated in Fig. 1 and the details discussed in this section.

We start again by looking at the derivative of the unitary evolution with respect to the parameters $\theta$. For every parameter, we can find a so-called effective generator $\mathcal{H}_j$ obeying

$$\frac{\partial}{\partial \theta_j} U(t_1, t_0) = -i U(t_1, t_0) \mathcal{H}_j. \tag{14}$$

$\mathcal{H}_j$ is called the "effective generator" because it resembles the derivative of a gate with a simple generator $\frac{\partial}{\partial \theta} e^{-i\theta G} = -i e^{-i\theta G} G$. We stress that the effective generators $\{\mathcal{H}_j\}$ are not to be confused with the pulse generators $\{H_j\}$ in Eq. (1), as they are generally different and only coincide in special cases[1].

The effective generator enables us to construct parameter-shift rules that we can execute on quantum hardware. As a first step, we use a differentiable ODE solver to obtain both $U(t_1, t_0)$ and its gradient $\nabla U(t_1, t_0)$. With those ingredients, we can construct the effective generator via Eq. (14) in terms of

$$\mathcal{H}_j = i U^\dagger(t_1, t_0) \frac{\partial}{\partial \theta_j} U(t_1, t_0). \tag{15}$$

Utilizing the general parameter-shift rules [8] at this point would require computing the eigenvalues of $\mathcal{H}_j$ and being able to execute $e^{-ix\mathcal{H}_j}$ with the appropriate shifts on the quantum computer. In particular, the latter is often not practical as it requires a decomposition of the unitary $e^{-ix\mathcal{H}_j}$ in terms of native gates on the quantum computer.

Instead, we decompose each effective generator into a basis of operators $\{P_j\}$ that generate rotations $e^{-i\frac{x}{2} P_j}$ that the quantum computer can execute. We choose Pauli words and write the decomposition

$$\mathcal{H}_j = \sum_\ell \omega_\ell^{(j)} P_\ell \tag{16}$$

via $\omega_\ell^{(j)} = \frac{\text{tr}[P_\ell \mathcal{H}_j]}{2^n}$. Overall, we have

---

[1] For example, when all $\{H_j\}$ mutually commute, or $\theta$ multiplies globally a sum of the pulse generators, $H(\theta, t) = \theta \sum_j f_j(t) H_j$.

$$\begin{aligned}
\frac{\partial \mathcal{L}}{\partial \theta_j} &= \langle\psi_0| \left[ U(t_1,t_0)^\dagger H_{\text{obj}} U(t_1,t_0), -i\mathcal{H}_j \right] |\psi_0\rangle \\
&= \sum_\ell 2\omega_\ell^{(j)} \langle\psi_0| \left[ U(t_1,t_0)^\dagger H_{\text{obj}} U(t_1,t_0), -\frac{i}{2} P_\ell \right] |\psi_0\rangle \\
&= \sum_\ell 2\omega_\ell^{(j)} \frac{d}{dx} \left[ \langle\psi_0| e^{i\frac{x}{2} P_\ell} U(t_1,t_0)^\dagger H^{\text{obj}} U(t_1,t_0) e^{-i\frac{x}{2} P_\ell} |\psi_0\rangle \right]_{x=0}.
\end{aligned} \quad (17)$$

Here, we introduced $x$ as a placeholder variable to identify the commutator as the derivative of a shifted expectation value

$$\mathtt{L}_\ell(x) = \langle\psi_0| e^{i\frac{x}{2} P_\ell} U(t_1,t_0)^\dagger H^{\text{obj}} U(t_1,t_0) e^{-i\frac{x}{2} P_\ell} |\psi_0\rangle. \quad (18)$$

With the two-term parameter-shift rule for Pauli words, $\frac{d}{dx}\mathtt{L}_\ell(x) = \frac{1}{2}\left(\mathtt{L}_\ell\left(x+\frac{\pi}{2}\right) - \mathtt{L}_\ell\left(x-\frac{\pi}{2}\right)\right)$, we can compute gradients on quantum hardware with

$$\frac{\partial \mathcal{L}}{\partial \theta_j} = \sum_\ell \omega_\ell^{(j)} \left( \mathtt{L}_\ell\left(\frac{\pi}{2}\right) - \mathtt{L}_\ell\left(-\frac{\pi}{2}\right) \right). \quad (19)$$

The derivation follows analogously for the case that $U(t_1,t_0)$ is only a part of a larger circuit, i.e., $V_1 U(t_1,t_0) V_0$ for some unitaries $V_1$ and $V_0$. Then, the shifted expectation values are given by

$$\mathtt{L}_\ell(x) = \langle\psi_0| V_0^\dagger e^{i\frac{x}{2} P_\ell} U(t_1,t_0)^\dagger V_1^\dagger H^{\text{obj}} V_1 U(t_1,t_0) e^{-i\frac{x}{2} P_\ell} V_0 |\psi_0\rangle. \quad (20)$$

Let us now look at the required quantum resources to compute gradients with ODEgen using Pauli words for the decomposition. The maximum number of expectation values $\mathtt{L}_\ell(\pm\frac{\pi}{2})$ that need to be executed is given by the dimension $d_{\text{DLA}}$ of the dynamical Lie algebra (DLA) [24] of the pulse generators. Take for example a pulse with generators $X_0$, $X_1$ and $Z_0 Z_1$. The DLA is spanned by $\{X_0, X_1, Z_0 Z_1, Y_0 Y_1, Y_0 Z_1, Z_0 Y_1\}$ and thus has a dimension of $d_{\text{DLA}} = 6$. Hence, the decomposition Eq. (16) has at most $d_{\text{DLA}}$ non-zero coefficients $\omega_\ell^{(j)}$. Note that we need not re-compute expectation values $\mathtt{L}_\ell(\pm\frac{\pi}{2})$ for different parameters $j$. In particular, we can cache measured values from the first occurrences of each $\mathtt{L}_\ell(\pm\frac{\pi}{2})$, and re-use them for subsequent effective generators.

The required quantum resources for ODEgen are thus in general

$$\mathcal{R}_{\text{ODEgen}} \leq 2 d_{\text{DLA}} \quad (21)$$

for a basis of Pauli words.

Depending on the choice of parameters $\theta_j$, some coefficients $\omega_\ell^{(j)}$ may happen to be zero. If they are consistently zero for all indices $j$, we can further reduce the number of expectation values. Beyond that, to additionally save resources we can choose to neglect contributions whose amplitudes are smaller than some user-defined threshold `atol`. This is all determined in classical preprocessing before executing the expectation values on quantum hardware. We call the number that counts the remaining terms $|\ell|$, leading to

$$\mathcal{R}_{\text{ODEgen}} = 2|\ell| \leq 2 d_{\text{DLA}} \quad (22)$$

expectation values overall.

Because we are using a classical ODE solver to obtain the effective generators that inform the parameter-shift rule terms, we are restricted to pulse gates that act on few qubits or realize a low-dimensional dynamical Lie algebra. This is in line with common practice in quantum algorithms to construct sufficiently expressive ansätze from smaller building blocks. Further, in many realistic scenarios on contemporary hardware like superconducting qubits, that is naturally the case due to the way interactions are implemented, as we will discuss in the following section.

## 3. Experiments with transmon Hamiltonians

Transmon qubit platforms can roughly be divided into fixed-frequency and tunable-frequency systems [1][25]. We focus on fixed-frequency systems due to their popularity with modern hardware vendors that provide access on the pulse level [26, 27, 20]. These platforms allow faithful single and two-qubit control due to their weak (fixed) couplings that have little effect when qubits are driven on resonance and enable two-qubit entangling gates when driven at cross-resonance [28, 29, 26] (i.e., driving one qubit at its neighboring qubit's frequency; see discussion around Eq. (26) below).

For such a coupled transmon Hamiltonian $H^{\mathcal{T}} = H^{\mathcal{T}}_{\text{drift}} + H^{\mathcal{T}}_{\text{drive}}$, the drift term is given by [1]

$$H^{\mathcal{T}}_{\text{drift}} = -\sum_q \frac{\omega_q}{2} Z_q + \sum_{q,p \in \mathcal{C}} J_{qp}(X_q X_p + Y_q Y_p), \tag{23}$$

with qubit frequencies $\omega_q$ and coupling strengths $J_{qp}$ for couplings $\mathcal{C} = \{(q,p)\}$ given by the device topology. The drive term

$$H^{\mathcal{T}}_{\text{drive}} = \sum_q \Omega_q(\theta, t) Y_q \tag{24}$$

has pulse envelopes $\Omega_q(\theta, t) = \Omega_q \, \mathbf{Re}\!\left(e^{i\nu_q t} u_q(\theta, t)\right)$ that are determined by the complex functions $|u_q(\theta, t)| \leq 1$, the maximum amplitudes $\Omega_q$, and the drive frequencies $\nu_q$. Note that one can encounter a different definition of the drive in terms of $\Omega_q(\theta, t) = \Omega_q v_q(\theta, t) \sin(\nu_q t + \varphi_q)$ with the real-valued functions $|v_q(\theta, t)| \leq 1$ and phases $\varphi_q$. By allowing $\varphi_q$ to vary in time, those two definitions are equivalent and we will use either interchangeably, depending on what is more convenient for the given context.

The physical parameters for the following experiments vary. Specific values can be found directly from [30].

### 3.1. Transmon physics

In the following section, we take a little detour into transmon physics, providing further details about the mechanism to perform single and two-qubit gates on these platforms. While doing so, we see that interactions between qubits are effectively reduced to pulses on few qubits, therefore motivating the use of ODEgen, whose main caveat is the restriction to low-dimensional subsystems. Readers already familiar with transmon physics may directly fast-forward to Sec. 3.2.

With transmon qubits, we can perform arbitrary single-qubit rotations in the X-Y-plane of the Bloch sphere, whereas rotations along the Z-axis are performed via virtual Z-gates [31]. The rotation axis in the X-Y-plane is set by the phase $\varphi_q$ of the drive. To understand this analytically, it is useful to transform the single-qubit transmon Hamiltonian $H^{\mathcal{T}} = -\frac{\omega_q}{2} Z_q + \Omega_q v(t) \sin(\nu_q t + \varphi_q) Y_q$ from the lab frame, where quantum computation occurs, into a rotating frame, the so-called qubit frame. This is achieved by transforming the time-evolved state via $R = e^{-i\frac{\omega_q t}{2} Z}$, leading to a new effective Hamiltonian in the qubit frame under which this state evolves, $\widetilde{H}^{\mathcal{T}} = i\dot{R}R^\dagger + RH^{\mathcal{T}}R^\dagger$.

Upon performing the rotating-wave approximation (see section IV D in [1]), this leads to a drive Hamiltonian in the qubit frame at resonance $\nu_q = \omega_q$ of

$$\widetilde{H}^{\mathcal{T}} = -\frac{\Omega_q}{2} v(\theta, t) \big(\cos(\varphi_q) X_q + \sin(\varphi_q) Y_q\big), \tag{25}$$

such that for $\varphi_q = 0$ / $\frac{\pi}{2}$ we realize a $X$ / $Y$ rotation in the qubit frame. Another way to think of this is that in the lab frame, the qubit precesses around the $Z$-axis. Driving the qubit on resonance additionally induces so-called Rabi-oscillations between the $|0\rangle$ and $|1\rangle$ state. Together, this leads to an overall spiral movement down the Bloch sphere when starting from the north pole ($|0\rangle$). The strength of the drive sets the relative angular velocity to the constant $Z$-precession, and the phase shifts the relative starting (and end) point of the $Z$-rotation, thereby setting the rotation axis as seen in the lab frame in Fig. 2.

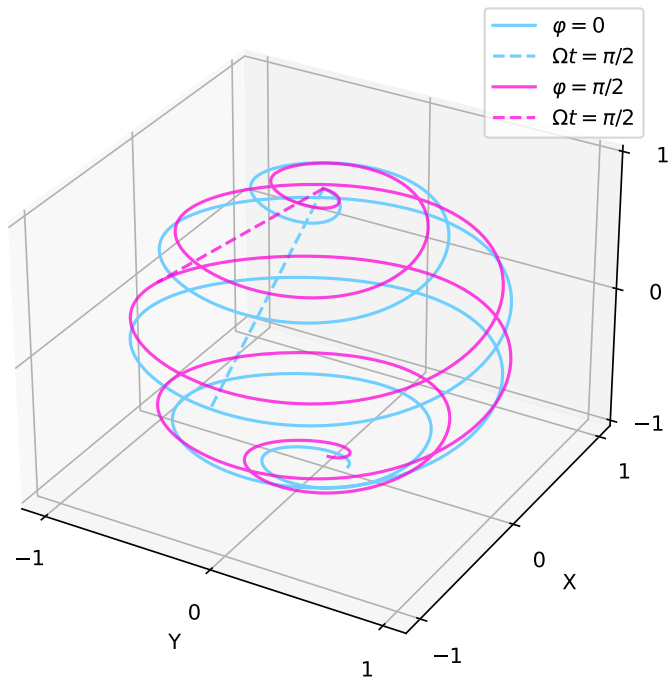

Fig. 2: Illustration of single-qubit driving on a Bloch sphere in the lab frame, starting from the north pole. A constant drive of $\Omega = 0.5\,\text{GHz}$ is applied on resonance $\nu = \omega = 5\,\text{GHz}$ over 1 ns. A constant precession around the $Z$-axis with the qubit's frequency $\omega$ is superimposed by the $Y$-rotation of the drive. Setting the phase $\varphi$ delays the $Z$-precession, effectively changing the rotation axis of the logical operation. The dotted lines indicate the logical operation of a $\frac{\pi}{2}$ drive achieved by $\Omega t = \frac{\pi}{2}$ with $\varphi = 0$ ($\varphi = \frac{\pi}{2}$), corresponding to a $\frac{\pi}{2}$ rotation around the $X$-axis ($Y$-axis). We provide a notebook to toy around with the parameters in [30].

With only single-qubit drives available, entanglement-generating gates are performed by cross-resonance driving [32, 26, 33, 34]. That is, driving one of two coupled transmon qubits at the resonance frequency of the other. In particular, driving qubit 0 at the resonance frequency of qubit 1, we obtain to first order

$$\widetilde{H}^{\mathcal{T}} \propto \frac{\Omega_0 J_{01}}{|\omega_1 - \omega_0|} (\cos(\varphi_0) Z_0 X_1 + \sin(\varphi_0) Z_0 Y_1) \tag{26}$$

in the rotating-qubit frame with rotating-wave approximations [32]. Thus, by setting $\varphi_0 = 0$ we can realize a ZX-rotation, which, in combination with single-qubit rotations, can generate a CNOT,

$$\text{CNOT} = \text{RZ}_0\left(\frac{\pi}{2}\right)\text{RX}_1\left(\frac{\pi}{2}\right)\text{ZX}_{01}\left(-\frac{\pi}{2}\right), \tag{27}$$

where $\text{RZ}_q(x) := e^{-i\frac{x}{2}Z_q}$, $\text{RX}_q(x) := e^{-i\frac{x}{2}X_q}$, and $\text{ZX}_{qp}(x) := e^{-i\frac{x}{2}Z_q X_p}$.

Note that Eq. (26) is only approximately correct with additional contributions occurring in the qubit frame. Those additional contributions can be derived and mitigated via echoed cross-resonance [33, 34]. For our numerical experiments, we are not using effective descriptions in the qubit frame, but look at the system in the lab frame (Eq. (23) and Eq. (24)).

The drive frequency of a pulse typically cannot be altered continuously, leading to the consequence that complicated multi-qubit gates need to be decomposed into a series of single-qubit on-resonance and two-qubit cross-resonance pulses. Aside from technical restrictions to continuously alter the drive frequency, it is not advisable to do so in the first place because it leads to loss landscapes with many narrow minima, see Fig. 3. This makes *any* optimization procedure with trainable drive frequency numerically difficult.

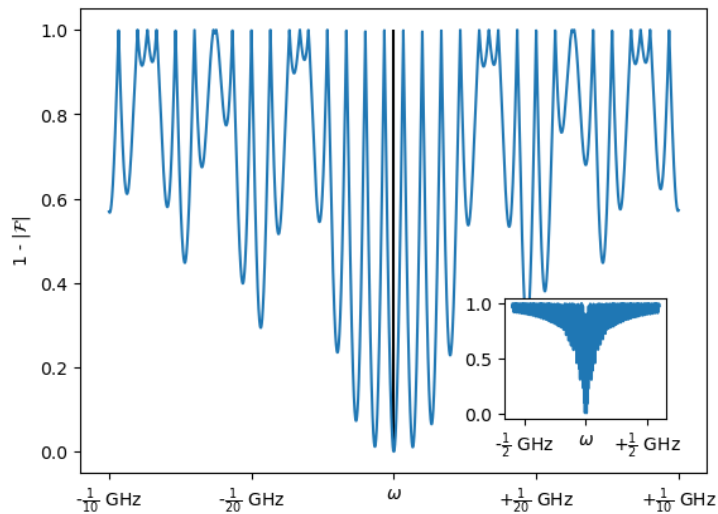

Fig. 3: After optimizing amplitude and phase of a single-qubit pulse gate to realize an $X$ gate, we vary the drive frequency $\nu$ around the qubit frequency $\omega$ and look at the infidelity $1 - |\mathcal{F}|$ with $\mathcal{F} = \text{tr}\left[U(\nu)^\dagger X\right]$. We find a loss landscape with many steep local minima and an overall trend towards maximum infidelity off-resonance. This makes the drive frequency very hard to optimize, and we usually instead build fixed ansätze with only resonant and cross-resonant drives.

Overall, we find that qubits are typically driven in only two modes: on resonance for single-qubit gates or on cross-resonance with neighboring connected qubits for entanglement generating gates.

This means that the restriction of ODEgen to few-qubit gates is not a problem in this scenario as we are always dealing effectively with single or few-qubit evolutions. In systems with tunable transmon qubits, driving on resonance is done in the same manner. The equivalent of driving on cross-resonance is tuning the qubit to match the frequency of the coupled, neighboring, qubit. Thus, also for tunable frequency transmons, interactions typically only occur in controlled, low-dimensional subsystems.

## 3.2. SPS gradients with transmon Hamiltonians

We now investigate the utility of the stochastic parameter-shift rule with realistic resources. In particular, we want to investigate the effect of the intrinsic stochasticity of the Monte-Carlo integration, Eq. (12). We use the transmon pulse Hamiltonian, Eq. (23) and Eq. (24), for a system of two qubits, and the pulse parametrization from [10]. For a fair comparison, we use the same physical parameters provided therein. Just like in [10], the pulses are parametrized by a sum of Legendre polynomials $P_\ell(t)$,

$$u_q(\theta, t) = \mathcal{N}\left(\sum_{\ell=0}^{d} \theta_\ell^q P_\ell\left(\frac{2t}{T} - 1\right)\right), \quad (28)$$

with $d = 4$. Each envelope has $d + 1$ complex parameters $\{\theta_\ell^q\}_{\ell=0}^{d}$, leading to $2(d+1)$ real parameters for each of the $n = 2$ qubits. The envelopes are normalized by $\mathcal{N}(z) = \frac{1-e^{-|z|}}{1+e^{-|z|}} e^{i \arg(z)}$ to ensure $|u_q| \leq 1$.

We show the signal-to-noise ratio between the expected gradient $\mathbb{E}(\nabla \mathcal{L})$ and its standard deviation over 100 random samples of size $N_s$ in Fig. 4. We show the mean and 90th percentile over the $n2(d+1) = 20$ parameters of $\nabla \mathcal{L}$. As an objective function, we take the expectation value of a tapered 2-qubit Hamiltonian of the HeH$^+$ molecule at 1.5Å [35]. Other 2-qubit Hamiltonians, such as the tapered $H_2$ molecule or ones with different bond lengths, lead to qualitatively the same results (we provide the code to produce Fig. 4, and to explore different Hamiltonians, parametrizations, and hyperparameters in [30]). We see that for realistic sample sizes, estimation of the gradient is heavily dominated by noise. We will see in the following section how this hinders effective training.

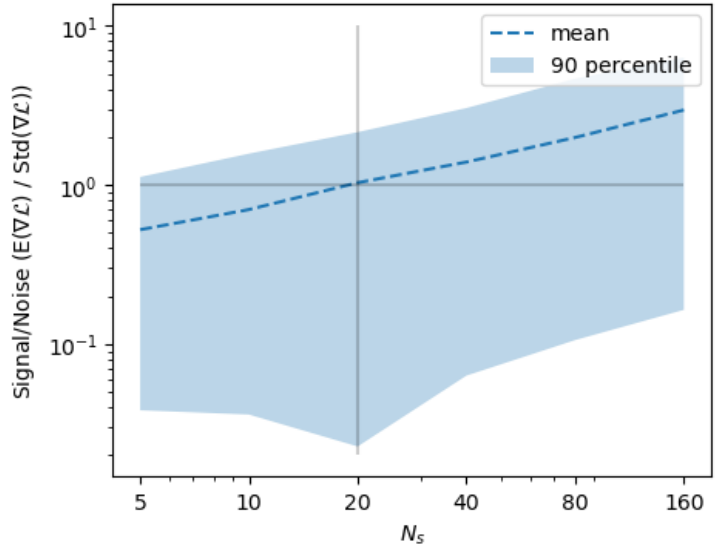

Fig. 4: Signal-to-noise ratio $\frac{\mathbb{E}(\nabla \mathcal{L})}{\text{Std}(\nabla \mathcal{L})}$ for gradients estimated with SPS as a function of the number of samples $N_s$ over the duration of the pulse. The parameters of Eq. (28) are evaluated at a fixed set of random parameters drawn from a Gaussian distribution centred around 0 with variance 1. We see the central limit theorem with the variance decreasing as $\frac{1}{\sqrt{N}}$, but that for practical sample sizes $N_s \leq 20$, the gradient estimator is highly dominated by noise. The dip at $N_s$ is a statistical fluctuation. Full details can be found in [30].

### 3.3. VQE experiments with transmon Hamiltonians

We run VQE experiments as described in [10] for the two-qubit $H_2$ and HeH$^+$ molecular Hamiltonians from [35] with the `adam` optimizer and learning rate set to 0.02 for 100 epochs. We compare the performance of the same VQE experiment using the analytic gradient obtained by ODEgen and the Monte Carlo approximation of the SPS in Fig. 5. For both cases, we choose the same 100 random initializations taken from a Gaussian distribution centered around 0 with variance 1.

It is worth noting that in this case the quantum resources to estimate the gradient, $\mathcal{R}_{\text{ODEgen}} = 30$, are lower than for the SPS method, which yields $\mathcal{R}_{\text{SPS}} = \{32, 80\}$ for $N_s = \{8, 20\}$ samples, respectively. Note that we can further reduce the resources of ODEgen by discarding all contributions $|\omega_\ell^j| \leq 1$ to the effective generator decomposition in Eq. (16). The savings depend on the set of parameters in the optimization procedure. We chose a case such that the optimization still works well, while savings are significant. For the employed threshold of atol = 1, we compute $\mathcal{R}_{\text{ODEgen}}^{\text{atol}=1} = 18.4 \pm 0.9$ expectation values on average per gradient evaluation during one optimization round of 100 epochs.

Overall, we find that we can reach lower energies with ODEgen while at the same time using fewer quantum resources. With the stochastic parameter-shift rule, it is unlikely to reach chemical accuracy ($10^{-3}$) for this and other tested examples (see [30]).

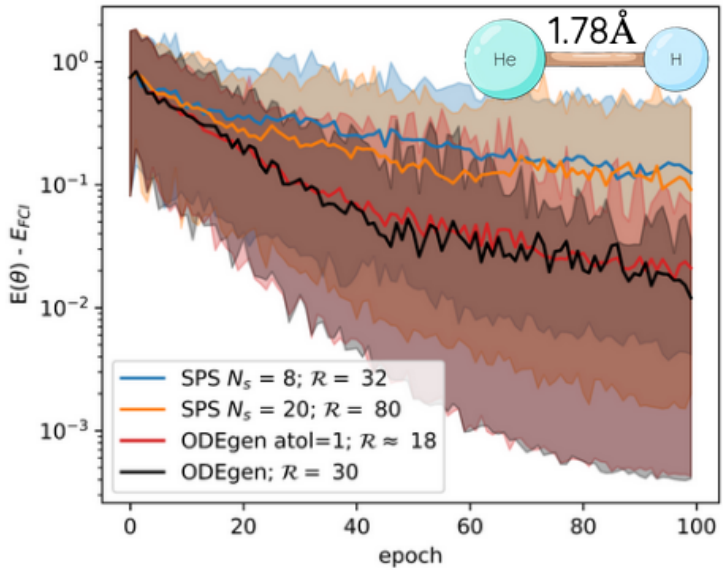

Fig. 5: Demonstration that with ODEgen we get better results while using fewer quantum resources, compared to using SPS for gradient evaluation. Running a VQE algorithm for HeH$^+$ (at 1.78Å in STO-3G basis, tapered to 2 spin orbitals) with parametrization as in Eq. (28). In both cases, we start from *the same* random parameters, initialized from a Gaussian distribution around 0 with variance 1. The lines show the mean value over 100 VQE runs, with the shaded regions showing the 90th percentile. (Blue and orange lines) Results with SPS gradients for $N_s = \{8, 20\}$ samples for the Monte Carlo integration. We further have $N_g = 2$ drive generators ($Y_0$ and $Y_1$) and $R = 1$ unique eigenvalue differences. This corresponds to $\mathcal{R}_{\text{SPS}} = \{32, 80\}$ expectation values on the device to estimate one gradient. (Red and black lines) Results with ODEgen with all $|\ell| = 4^2 - 1 = 15$ terms s.t. $\mathcal{R}_{\text{ODEgen}} = 30$. Further discarding all contributions $|\omega_\ell^j| \leq$ atol = 1 leads to $\mathcal{R}_{\text{ODEgen}} \approx 18$ on average (as it depends on the decomposition for the given parameters) with similar results. Full details of the numerics can be found in [30].

In the previous simulation, we used one single-pulse gate as the entire circuit, as was done in [10, 5]. For more complex systems, we need an ansatz combining on-resonance and cross-resonance drives. We know that a combination of two resonant single-qubit drives and a two-qubit cross-resonance drive can express a CNOT gate with single-qubit rotations. We use the ansatz described in Fig. 6(a). This is an echoed cross-resonance ansatz [33, 36] sandwiched by single-qubit resonant drives. Each gate has a trainable amplitude and phase with 10 time segments per gate, as well as a fixed time of $T = 20$ns and 100ns for resonant and cross-resonant driving, respectively.

One potential application of using pulse gradients on hardware is to replace parts of an existing circuit with an appropriate pulse gate. The goal is not to optimize the pulse gate to match the removed gates it is replacing; instead, we here directly optimize the cost function of the variational quantum algorithm as we would have with an ansatz consisting of all digital gates.

We show this exemplarily for the $H_4$ molecule at 1Å equidistant separations in the STO-3G basis with 8 spin orbitals (qubits) after a Jordan-Wigner transformation. We take both the molecular Hamiltonian and a standardized, pre-optimized VQE ansatz from [35] with a structure of single and double excitations [37, 38]. We take this fixed structure and replace parts of it with our pulse gate ansatz above. In particular, we replace a single CNOT with all its surrounding single-qubit drives. We illustrate the altered circuit in Fig. 6(b) and plot the training curve in Fig. 6(c).

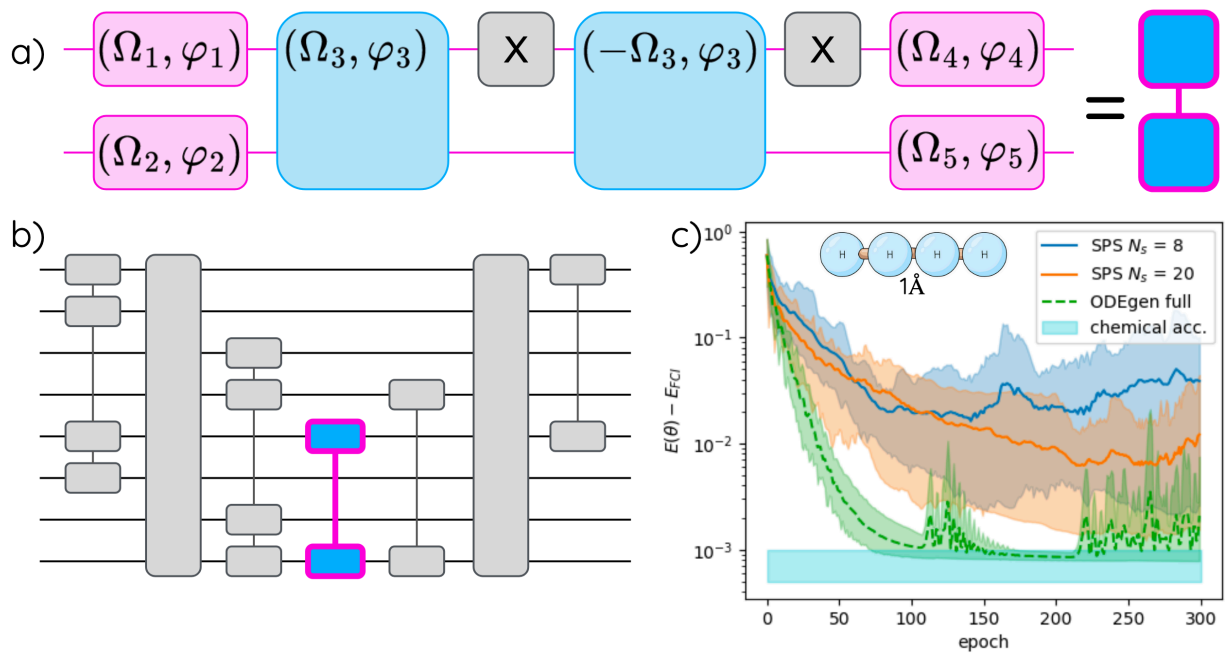

Fig. 6: VQE experiment using the $H_4$ molecule at 1Å and 180° in STO-3G, leading to 8 spin orbitals (qubits). Full details available in [30]. a) Two-qubit pulse gate ansatz consisting of two single drives (pink, 20ns), a cross-resonant drive on the first qubit (blue, 100ns), followed by an X gate, the same cross-resonant pulse but with opposite amplitude, followed by another X gate and again two single resonant drives b) Fixed singles and doubles ansatz, but part of it is replaced with the pulse gate ansatz described in a). c) Optimizing the pulse gate while keeping the rest fixed. We show the average (lines) and 90th percentile (shaded regions) over the same 16 random initializations using SPS and ODEgen. Just like in Fig. 5, we have quantum resources $\mathcal{R}_{\text{SPS}} = \{32, 80\}$ and $\mathcal{R}_{\text{ODEgen}} = 30$. The optimization using ODEgen converges to chemical accuracy ($10^{-3}$) within 100 epochs, while the SPS optimization struggles to reach it within 300 epochs. The VQE energy $-2.16553$ obtained from the original circuit is slightly above the minimal energy obtained with the pulse gate $-2.16561$.

## 3.4. Experiments on hardware

We perform a small-scale VQE example over $N_{it} = 14$ iterations fully on hardware using ODEgen for gradient estimation of the pulse gate. The results are displayed in Fig. 7.

As an objective function, we use $H_{obj} = Z$, i.e., $\mathcal{L} = \langle Z \rangle$, and a circuit that consists solely of a constant pulse with trainable amplitude $\Omega$ over 20ns. We run our experiment through `PennyLane` and Amazon Braket in the cloud on `Lucy`, a fixed-frequency transmon quantum computer by *Oxford Quantum Circuits* (OQC). The employed transmon qubit has a resonance frequency of $\omega = 4.472 \times 2\pi$ GHz and we measure an attenuation $\eta \approx 0.14$ between the output voltage $V_0$ on the arbitrary waveform generator that we set on the device, versus the actual voltage $V_{device} = \eta V_0$ that the transmon qubits receive. This allows us to perform better classical simulations for comparison; see Fig. 7 b). Full details on how we measure $\eta$ can be found in [30].

Since we only want to minimize $\mathcal{L} = \langle Z \rangle$, we do not need to compute all three $X$, $Y$, and $Z$ contributions to the effective generator in the drive's dynamical Lie algebra. It suffices to just take one of $X$ and $Y$ contributions. For this, we choose the constant angle $\varphi$ such that we get mostly $X$ contributions and compute the gradients accordingly.

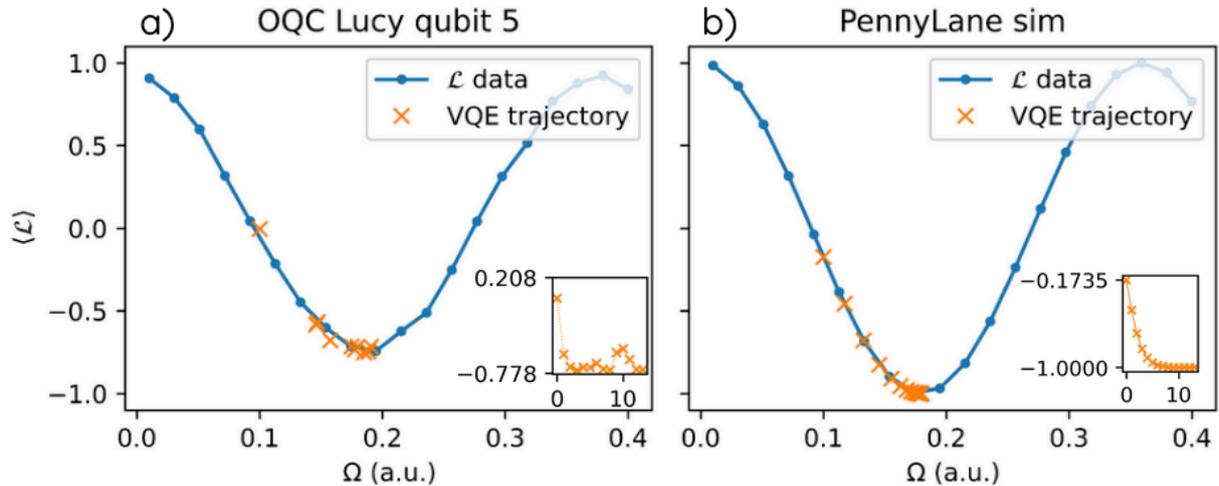

Fig. 7: Loss landscape of $\mathcal{L} = \langle Z \rangle$ after evolving with a constant pulse amplitude $\Omega$ over 20ns. We map out the loss landscape and see the Rabi oscillations, to which we fit a sinusoid. We perform a VQE optimization of the amplitude starting from $\Omega = 0.1$ with vanilla gradient descent using ODEgen to find the minimum and plot the trajectory in the loss landscape. The insets show the optimization by epochs. a) Performance of qubit 5 (zero-based numbering) on OQC Lucy with learning rate 0.01 over $N_{it} = 14$ epochs. b) Performance in simulation after calibrating the attenuation $\eta \approx 0.14$ of the device, with a learning rate of 0.001 over $N_{it} = 14$ epochs. The exact details can be found in [30].

By restricting ourselves to only one element in the Lie algebra ($\ell = 1$) and only one term $Z$ to measure, we have a total of $\mathcal{R} = N_{it}(2\ell + 1)$ expectation values. With the employed $N_{shots} = 1000$ and current market rates this yields a cost of \$27.3 for Fig. 7(a). That is not including calibrations ($\mathcal{O}(10)$ expectation values) and tuning hyperparameters ($\mathcal{O}(10)$ settings using $\mathcal{O}(10-100)$ iterations), elevating the overall endeavor to $\mathcal{O}(\$100)$.

The employed qubit has a limited readout fidelity of 90.75% at the time of conducting our experiments. This quantity is provided by the vendor and is significantly higher than our observations in the Rabi oscillations with a fitted magnitude of 0.83 (see Fig. 7(a)). One possible source of errors is excitations to higher levels. Though the chosen amplitudes are relatively low and should not cause higher-level excitations, the shape of the constant pulse leads to sharp derivatives at the start and end of the pulse, potentially leading to higher level excitations [39, 40]. Another source of noise may come from coherent and classical cross-talk that is currently not explicitly taken into account by our model [36]. To make gradient-based optimization on hardware viable we will need to improve both the classical modeling of the hardware, as well as the fidelity of the hardware itself.

## 4. Conclusion

We derived a new analytic method for computing gradients of pulse programs on quantum computers by leveraging differentiable ODE solvers in `jax`. We saw that for many relevant systems, the assumption of low-dimensional subsystems is valid and further demonstrated advantages of using ODEgen over the stochastic parameter-shift rule in that it can reach lower minima in variational quantum algorithms with similar or fewer quantum resources.

We make all numerical experiments, as well as runs on real quantum hardware shown in this paper, accessible and reproducible in [30]. An open-source implementation of ODEgen has been included as part of the recently developed pulse-programming module of `PennyLane` [21].

Both the stochastic parameter-shift rule and ODEgen are valid for all pulses of the form Eq. (1). We specifically looked at transmon qubits with fixed coupling strengths and all-microwave driving. Most of the results are transferable to other quantum computing platforms that are based on electromagnetic driving of qubits, such as, for example, ion trap and neutral atom systems. We note, however, that for as long as neutral atom quantum computers cannot perform digital gates, neither the stochastic parameter-shift rule nor ODEgen are viable to compute gradients. That is because, in either case, we require the execution of digital gates $e^{-i(\pm\frac{\pi}{4}H_j)}$.

As a proof of concept, we performed a simple VQE example fully on hardware. There is still a big gap between running variational quantum algorithms in simulation and on real hardware. If we want to realistically run such algorithms on interesting problem sizes we need to improve the limited availability of devices, high dollar cost per quantum resource $\mathcal{R}$, qubit quality, noise levels, classical modeling, and algorithms. For the latter, we contributed a new analytic gradient technique that requires fewer quantum resources $\mathcal{R}$ for relevant problems.

## Acknowledgements

Tom Bromley, Lillian Frederiksen, Korbinian Kottmann, Albert Mitjans Coma, Mudit Pandey, Filippo Vicentini, and David Wierichs have contributed to the open-source implementation of all pulse programming functionality in `PennyLane`. The authors would further like to thank Lillian Frederiksen, Jean-Christophe Jaskula, Filippo Vicentini, and David Wierichs for fruitful discussions about the contents of this paper.

## Bibliography

[1] P. Krantz, M. Kjaergaard, et al., "A quantum engineer's guide to superconducting qubits," *Appl. Phys. Rev.*, vol. 6, no. 2, Jun. 2019. https://doi.org/10.1063/1.5089550

[2] C. D. Bruzewicz, J. Chiaverini, R. McConnell, and J. M. Sage, "Trapped-ion quantum computing: progress and challenges," *Appl. Phys. Rev.*, vol. 6, no. 2, May 2019. https://doi.org/10.1063/1.5088164

[3] L. Henriet, L. Beguin, et al., "Quantum computing with neutral atoms," *Quantum*, vol. 4, p. 327, Sep. 2020. https://doi.org/10.22331/q-2020-09-21-327

[4] L. Clinton, J. Bausch, and T. Cubitt, "Hamiltonian simulation algorithms for near-term quantum hardware," *Nature Commun.*, vol. 12, no. 1, Aug. 2021. https://doi.org/10.1038/s41467-021-25196-0

[5] O. R. Meitei, B. T. Gard, et al., "Gate-free state preparation for fast variational quantum eigensolver simulations: ctrl-VQE," 2021. https://arxiv.org/abs/2008.04302

[6] A. Asthana, C. Liu, et al., "Minimizing state preparation times in pulse-level variational molecular simulations," 2022. https://arxiv.org/abs/2203.06818

[7] O. Kyriienko, and V. E. Elfving, "Generalized quantum circuit differentiation rules," *'Phys. Rev. A'*, vol. 104, no. 5, Nov. 2021. https://doi.org/10.1103/PhysRevA.104.052417

[8] D. Wierichs, J. Izaac, C. Wang, and C. Y.-Y. Lin, "General parameter-shift rules for quantum gradients," *Quantum*, vol. 6, p. 677, Mar. 2022. https://doi.org/10.22331/q-2022-03-30-677

[9] L. Banchi, and G. E. Crooks, "Measuring Analytic Gradients of General Quantum Evolution with the Stochastic Parameter Shift Rule," *Quantum*, vol. 5, p. 386, Jan. 2021. https://doi.org/10.22331/q-2021-01-25-386

[10] J. Leng, Y. Peng, Y.-L. Qiao, M. Lin, and X. Wu, "Differentiable analog quantum computing for optimization and control," 2022. https://arxiv.org/abs/2210.15812

[11] R. Wiersema, D. Lewis, D. Wierichs, J. Carrasquilla, and N. Killoran, "Here comes the SU(N): multivariate quantum gates and gradients," 2023. https://arxiv.org/abs/2303.11355

[12] J. Dormand, and P. Prince, "A family of embedded Runge-Kutta formulae," *J. Comput. Appl. Math.*, vol. 6, no. 1, pp. 19–26, 1980. https://doi.org/10.1016/0771-050X(80)90013-3

[13] R. T. Q. Chen, Y. Rubanova, J. Bettencourt, and D. Duvenaud, "Neural ordinary differential equations," 2019. https://arxiv.org/abs/1806.07366

[14] N. Khaneja, T. Reiss, C. Kehlet, T. Schulte-Herbrüggen, and S. J. Glaser, "Optimal control of coupled spin dynamics: design of NMR pulse sequences by gradient ascent algorithms," *J. Magnetic Reson.*, vol. 172, no. 2, pp. 296–305, 2005. https://www.sciencedirect.com/science/article/pii/S1090780704003696

[15] P. de Fouquieres, S. Schirmer, S. Glaser, and I. Kuprov, "Second order gradient ascent pulse engineering," *J. Magnetic Reson.*, vol. 212, no. 2, pp. 412–417, Oct. 2011. https://doi.org/10.1016/j.jmr.2011.07.023

[16] P. Gokhale, Y. Ding, et al., "Partial compilation of variational algorithms for noisy intermediate-scale quantum machines," in *Proc. 52nd Annu. IEEE/ACM Int. Symp. Microarchitecture*, Oct. 2019. https://doi.org/10.1145/3352460.3358313

[17] M. H. Goerz, S. C. Carrasco, and V. S. Malinovsky, "Quantum optimal control via semi-automatic differentiation," *Quantum*, vol. 6, p. 871, Dec. 2022. https://doi.org/10.22331/q-2022-12-07-871

[18] M. Cerezo, A. Arrasmith, et al., "Variational quantum algorithms," *Nat. Rev. Phys.*, vol. 3, no. 9, pp. 625–644, Aug. 2021. https://doi.org/10.1038/s42254-021-00348-9

[19] K. Bharti, A. Cervera-Lierta, et al., "Noisy intermediate-scale quantum algorithms," *Rev. Mod. Phys.*, vol. 94, no. 1, Feb. 2022. https://doi.org/10.1103/revmodphys.94.015004

[20] J. Rahamim, T. Behrle, et al., "Double-sided coaxial circuit QED with out-of-plane wiring," *Appl. Phys. Lett.*, vol. 110, no. 22, May 2017. https://doi.org/10.1063/1.4984299

[21] V. Bergholm, J. Izaac, et al., "PennyLane: automatic differentiation of hybrid quantum-classical computations," 2022. https://arxiv.org/abs/1811.04968

[22] "Amazon Braket," 2023. https://aws.amazon.com/braket/

[23] J. Bradbury, R. Frostig, et al., "JAX: composable transformations of Python+NumPy programs," 2018. http://github.com/google/jax

[24] M. Larocca, P. Czarnik, et al., "Diagnosing Barren Plateaus with Tools from Quantum Optimal Control," *Quantum*, vol. 6, p. 824, Sep. 2022. https://doi.org/10.22331/q-2022-09-29-824

[25] A. Blais, A. L. Grimsmo, S. M. Girvin, and A. Wallraff, "Circuit quantum electrodynamics," *Rev. Mod. Phys.*, vol. 93, no. 2, May 2021. https://doi.org/10.1103/revmodphys.93.025005

[26] J. M. Chow, A. D. Corcoles, et al., "Simple all-microwave entangling gate for fixed-frequency superconducting qubits," *Phys. Rev. Lett.*, vol. 107, no. 8, Aug. 2011. https://doi.org/10.1103/physrevlett.107.080502

[27] T. Alexander, N. Kanazawa, et al., "Qiskit pulse: programming quantum computers through the cloud with pulses," *Quantum Sci. Technol.*, vol. 5, no. 4, p. 44006, Aug. 2020. https://doi.org/10.1088/2058-9565/aba404

[28] J. Koch, T. M. Yu, et al., "Charge-insensitive qubit design derived from the Cooper pair box," *'Phys. Rev. A'*, vol. 76, no. 4, Oct. 2007. https://doi.org/10.1103/physreva.76.042319

[29] J. A. Schreier, A. A. Houck, et al., "Suppressing charge noise decoherence in superconducting charge qubits," *Phys. Rev. B*, vol. 77, no. 18, May 2008. https://doi.org/10.1103/physrevb.77.180502

[30] K. Kottmann, "Evaluating analytic gradients of pulse programs on quantum computers," 2023. https://github.com/XanaduAI/Analytic_Pulse_Gradients

[31] D. C. McKay, C. J. Wood, S. Sheldon, J. M. Chow, and J. M. Gambetta, "Efficient Z-gates for quantum computing," *'Phys. Rev. A'*, vol. 96, no. 2, Aug. 2017. https://doi.org/10.1103/physreva.96.022330

[32] C. Rigetti, and M. Devoret, "Fully microwave-tunable universal gates in superconducting qubits with linear couplings and fixed transition frequencies," *Phys. Rev. B*, vol. 81, no. 13, p. 134507, Apr. 2010. https://link.aps.org/doi/10.1103/PhysRevB.81.134507

[33] S. Sheldon, E. Magesan, J. M. Chow, and J. M. Gambetta, "Procedure for systematically tuning up cross-talk in the cross-resonance gate," *'Phys. Rev. A'*, vol. 93, no. 6, p. 60302, Jun. 2016. https://link.aps.org/doi/10.1103/PhysRevA.93.060302

[34] E. Magesan, and J. M. Gambetta, "Effective Hamiltonian models of the cross-resonance gate," *'Phys. Rev. A'*, vol. 101, no. 5, May 2020. https://doi.org/10.1103/physreva.101.052308

[35] U. Azad, "PennyLane quantum datasets," 2023. https://pennylane.ai/datasets/

[36] A. Patterson, J. Rahamim, et al., "Calibration of a cross-resonance two-qubit gate between directly coupled transmons," *Phys. Rev. Appl.*, vol. 12, no. 6, p. 64013, Dec. 2019. https://link.aps.org/doi/10.1103/PhysRevApplied.12.064013

[37] G.-L. R. Anselmetti, D. Wierichs, C. Gogolin, and R. M. Parrish, "Local, expressive, quantum-number-preserving VQE ansätze for fermionic systems," *New J. Phys.*, vol. 23, no. 11, p. 113010, Nov. 2021. https://doi.org/10.1088/1367-2630/ac2cb3

[38] J. M. Arrazola, O. D. Matteo, et al., "Universal quantum circuits for quantum chemistry," *Quantum*, vol. 6, p. 742, Jun. 2022. https://doi.org/10.22331/q-2022-06-20-742

[39] F. Motzoi, J. M. Gambetta, P. Rebentrost, and F. K. Wilhelm, "Simple pulses for elimination of leakage in weakly nonlinear qubits," *Phys. Rev. Lett.*, vol. 103, no. 11, Sep. 2009. https://doi.org/10.1103/physrevlett.103.110501

[40] J. M. Gambetta, F. Motzoi, S. T. Merkel, and F. K. Wilhelm, "Analytic control methods for high-fidelity unitary operations in a weakly nonlinear oscillator," *'Phys. Rev. A'*, vol. 83, no. 1, Jan. 2011. https://doi.org/10.1103/physreva.83.012308